\documentstyle[12pt,psfig]{article}
\begin{document}

\title{The alpha decay of deformed superheavy elements.\thanks{
The work was partially sponsored by the State Committee for Scientific Research
under contract No. 2P 03B 115 19}}

\author{     Micha\l{} Kowal,  Zdzis\l{}aw \L{}ojewski           \\ 
        {\it Department of Theoretical Physics, Institute of Physics,}  \\ 
        {\it Maria Curie-Sk\l{}odowska University, 20-031 Lublin, Poland }}

%\date{}

\maketitle

\begin{abstract}
The interaction potential between the alpha particle and the deformed parent nucleus was used for
description of the decay of superheavy nuclei. It consists of centrifugal, nuclear and Coulomb
parts suitably modified for deformed nuclei. The significant effect of various shapes of barriers
obtained for  deformed parent nuclei on  calculated alpha half-life times were shown.
The finally calculated half-life times due to the spontaneous alpha decay for superheavy elements 
were compared with the results obtained from other models and the experimental data.
\end{abstract}

\section{Introduction}
Theoretical  calculations suggesting  existence of deformed superheavy nuclei 
from the areas of doubly magic $^{270}Hs$ were made by Sobiczewski et. al.\cite{PATYK91,2PATYK91}. 
Recently there have been calculated some collective properties of  nuclei such as: deformation energies
with the equilibrium shapes, energies of the first excited state $E_{2+}$ and the branching
ratios  between $\alpha$-decay to this state $2+$, to that to the ground state $0+$.
Possible puzzling comparison of the obtained  theoretical predictions with the experimental data
promotes  and motivates further studies of this problem. Therefore more extensive description of an 
$\alpha$ particle emission including additional degrees of freedom connected with deformation in 
ground state of emitting nucleus is necessary. Then different possition of $\alpha$ particle towards this
nucleus is possible.

It is commonly known that the  action integral giving probability of $\alpha$ particle  tunneling
depends strongly, among others, on the potential barrier shapes and even insignificant change of this 
barrier can change the obtained half-lives by the magnitude of the a few orders. Taking deformation  of
parent nucleus into account in our model can give more realistic results. 

The aim of the paper is calculation of half-lives due to spontaneous $\alpha$ decay for the deformed
superheavy nuclei which  is a predominant process for them. The barriers  were calculated using a
suitable model in which the total interaction potential ($V_{TOT}$) consists of the nuclear ($V_{N}$),
Coulomb ($V_{C}$) and centrifugal  barrier ($V_{L}$) modified for the case of 
existing the deformation of parent nucleus.
Chapter II includes brief description of all terms of the potential. Chapter III presents
main formulas relating to the $\alpha$ decay used in this paper. Chapter IV includes the discussion 
of the results and the influence on half-lives of the various factors. 
The last chapter includes summing up and conlusions. 

\section{Interaction potential}
\subsection{Shape parameterization in the exit channel}
The shape of nucleus surface area  $R_{P(\alpha)}(\theta_{P(\alpha)},\phi_{P(\alpha)})$ is
parameterized in the standard way in the spherical harmonic base which can be written in the vector
notation as follows:
\begin{equation}
\vec r_{P(\alpha)} =
\left\{
\begin{array}{ll}
R_{P(\alpha)}(\theta_{P(\alpha)},\phi_{P(\alpha)})\sin(\theta_{P(\alpha)})\cos(\phi_{P(\alpha)})\cr
R_{P(\alpha)}(\theta_{P(\alpha)},\phi_{P(\alpha)})\sin(\theta_{P(\alpha)})\sin(\phi_{P(\alpha)})\cr
R_{P(\alpha)}(\theta_{P(\alpha)},\phi_{P(\alpha)})\cos(\theta_{P(\alpha)})
\end{array}   
\right.
\label{geo}
\end{equation}
The index $P(\alpha)$ refers to parent nucleus or alpha particle adequately. 
In the reference system connected with the parent nucleus the centre of the emitted particle 
is parameterized in the spherical  system  ($R,\Theta,\Phi$):
\begin{equation}
\vec R =
\left\{
\begin{array}{ll}
R\sin(\Theta)\cos(\Phi)\cr
R\sin(\Theta)\sin(\Phi)\cr
R\cos(\Theta)
\end{array}   
\right.
\end{equation} 
Figure 1 shows geometry of this system where the shortest distance between the nucleus surface areas
of $\alpha$ particle and the parent nucleus is denoted by s, the distance between the mass centers by R  
and the distance between the infinitely small volume elements in the nucleus of emitter and emitted
particle by $r_{P\alpha}$.

Particle $\alpha$ is treated as a sphere of a radius  $R_{\alpha}=1.84 fm$. 
There were considered only axially symmetric deformed parent nuclei which results in the
simplification - $\phi_{P(\alpha)} $ = 0 and  $\Phi$ = 0.

\subsection{Coulomb potential}
The electrostatic interaction potential ($V_{C}$) between the parent nucleus with the proton ($Z_{P}$),
mass numbers ($A_{P}$) and the emitted $\alpha$  particle is given  by integral equation:
\begin{equation}
\label{T,P}
{V_{C}} = {2e{\int_{\nu_{P}} \frac{\rho_{P}(\vec r_{P})}{r_{P\alpha}} d\tau_{P}}} \,\,, 
\label{coul}
\end{equation}
where $e$ is the charge unit and $\rho_{P}(\vec r_{P})$ is a charge distribution with uniform density.
It is $\rho_{P}(\vec r_{P})=\frac{3Z_{P}e}{4 \pi R^{\ 3}_{CP}}$ inside the parent nucleus 
and $\rho_{P}(\vec r_{P})= 0$ outside the nucleus.
The {\it  equivalent  sharp radius} is taken as ${R_{CP}} = 1.15 A^{\frac{1}{3}}_{P}$. 
Integration over the  parent nucleus volume ($\nu_{P}$) in the coordinates defined by equation 
(\ref{geo})  is carried out.
In the outside area, when the distance between the mass centers is larger than the distances
corresponding to the contact configuration of the parent nucleus and the alpha particle
$R>r_{P}+r_{\alpha}$, expansion of distance reciprocity $r_{P\alpha}$ is of absolutely convergent
series form:
\begin{equation}
\frac{1}{r_{P\alpha}} = \frac{1}{2\sqrt{\pi}}\sum_{l_{P}}^{\infty}\frac{(1)^{l_{P}}}{R^{l_{P}+1}}
C_{l_{P}, 0}^{0, 0} \ Y_{l_{P}}^{0}(\Theta, \Phi) Y_{l_{P}}^{0}(\theta_{P}) \,\,. 
\label{exp}
\end{equation}
After substitution of the expansion into the formula (\ref{coul}) and making simple transformations we
obtain suitable multipole expansion for the Coulomb  interactions of deformed nuclei:
\begin{eqnarray}
V_{C} =  \frac{2 Z_{P}e^{2}}{R}+\frac{e}{\sqrt\pi}\sum_{l_{P} = 2}^{\infty} 
\frac{1}{R^{l_{P}+1}}
C_{l_{P},0}^{\ 0,0} 
\ Y_{l_{P}}^{0}(\Theta)Q_{l_{P}}^{0}   \,\,, 
\end{eqnarray} 
Here $C_{l_{P},0 }^{\ 0,0}$ are the Wigner coefficients. 
The required multipole moments $Q^{0}_{l_{P}}$ are calculated using the following 
definition \cite{HASSE88}:
\begin{equation}
Q^{0}_{l_{P}} = {\int_{\nu_{P}} \rho_{P}(\vec r_{P}) r_{P}^{l_{P}}
Y_{l_{P}}^{0}(\theta_{P}) d\tau_{P}} \,\,.
\end{equation}
here $Y_{l_{P}}^{0}(\theta_{P})$ are corresponding sherical harmonics.
Let us note that the first term in the presented multipole expansion corresponds 
to the case of interaction when both fragments are spherical. 

\subsection{Nuclear potential}
The nuclear part ($V_{N}$) was calculated using  proximity aproximation \cite{BLOCKI77}.
This hypothesis is based on the assumption, that the nuclear interaction takes place
between two infinitesimal volume elements located
closest for a given configuration, and additionally that the distance between 
them is of the order of a few fm.
The surfaces could be treated then as flat parrallel slabs and the proximity 
energy between two semi-infinite nuclear matters is of a simpler form: 
\begin{equation}
V_{N} = K \ \gamma \ \psi(s) \,\,,
\end{equation}
with the surface tension coefficient $\gamma = 0.9517(1-1.7826I^{2})MeV/fm^{2}$,
where $I=(N-Z)/A$.
It allows for division of nucleus-nucleus interactions into two independent factors: $K$ connected with
the shape of emitting nucleus surface area and the factor expressed by the universal function $\psi(s)$,
independent of the system geometry. The form of universal function given in \cite{MYERS00} is used in
this paper. 

Expansion of this approach for the deformed case consists in suitable redefining of the geometrical
factor. In the case of deformed nuclei it is the geometrical mean of {\it main reduced curvature radii} 
$k_{1}$ and $k_{2}$: 
\begin{equation}
K_{DEF}=2\pi \sqrt{k_{1} \cdot {k_{2}}} \,\,,
\end{equation}  
calculated on the nucleus surface area in the sites for which with the data R i $\Theta$   
the distance between surfaces is minimal $s = min \vert \vec R+\vec r_{\alpha}-\vec r_{P} \vert$.
To calculate the required curvature radii, we calculate fundamental 
forms of the first and the second orders 
of the nucleus surface in these  points.
Then we are able to calculate  the Gauss curvature and  the mean curvature 
and  finally we obtain the first $R_{1M(\alpha)}$ 
and second $R_{2M(\alpha)}$ main radius curvatures \cite{BRONSTEIN}.
The {\it main reduced curvature radius} of the first
and second orders $k_{1}$ and $k_{2}$ can be expressed by the relations:
\begin{equation}
k_{1} ={\frac{R_{1M}R_{1\alpha}}{R_{1M}+R_{1\alpha}}} ; \  k_{2}
={\frac{R_{2M}R_{2\alpha}}{R_{2M}+R_{2\alpha}}} \,\,. 
\end{equation}  
Moreover, within small deformations the factor  $K_{DEF}$ passes into the factor $K$ for the spherical
case $K_{DEF}({\beta_{l_{P}}=0})->K$.
 
\section{Theory of alpha decay}
The emission of alpha particle from the axially symmetric deformed 
superheavy nuclei is considered.
The  half-life is given as a function of the position  $\alpha$ particle:
\begin{equation}
T_{1/2} (\Theta) = \frac{ln2}{\lambda (\Theta)}  \,\,,
\end{equation} 
where $\lambda$ is the decay constant.
The $\alpha$-decay model for deformed superheavy elements presented here is based on the
assumption that forming of  $\alpha$ cluster in the inside of 
parent nucleus and passing over potential barrier are independent events.
It is also assumed that the emission of $\alpha$ particle does not affect
deformation of the nucleus from which it was emitted.
Within these approximations the $\alpha$-decay tunnelling probability
constant can be writen as:
\begin{equation}
\lambda (\Theta)= S \, \lambda_{G}(\Theta) \,\,,
\end{equation}
where S is the quantum mechanical probability forming of $\alpha$ particle in the parent nucleus 
({\it preformation  probability}). It is additionally assumed that the deformation and orientation
of parent nucleus do not have any influence on the {\it preformation  probability}.
By the  $\lambda_{G}(\Theta)$ we denoted the tunneling probability ({\it Gamow decay constant}), 
which is a function of the energy of the $\alpha$ particle and angle $\Theta$ showing the direction
of $\alpha$ emission.
All informnation about the many body structure is contained in the S and because of it 
we sometimes call  this constant the {\it spectroscopic factor}.
This factor can be calculated as the mean value of the suitable projection operator $\widehat{P}$
 between  parent nucleus wave function $\Phi_{P}$ lies in the open channel space 
$\Phi_{D} \otimes \Phi_{\alpha}$ spread 
by the daughter nucleus wave function $\Phi_{D}$ and $\alpha$ nucleus wave function $\Phi_{\alpha}$.   
\begin{equation}
S = \langle \Phi_{P} \vert \widehat{P} \vert \Phi_{P} \rangle \,\,.
\end{equation} 
This problem was invesigated by Blendowske, Fliessbach and Walliser \cite{FLIESS96} 
for different exotic spontaneous emissions of clusters and for  $\alpha$ particle needed here.
Exact numerical evaluation and fitting procedure lead to the semiempirical formula 
for {\it spectroscopic factor}.
In the  $\alpha$ particle case this factor  is equal to $S_{\alpha}=6.3 \cdot 10^{-3}$ \cite{FLIESS96}.
The favoured decays are considered here only.
The barrier penetration probability $\lambda_{G}$ as a funcion of the  angle can be
expressed:  
\begin{equation}
 \lambda_{G}(\Theta) = \frac{\omega(\Theta)}{2\pi} \  e^{I(\Theta)} \,\,,
\label{lmg}
\end{equation}
where $I(\Theta)$ is the action integral:
\begin{equation}
 I(\Theta) =  
 \int_{_{_{\! \! \! \! \! \! \! r_{en}(\Theta)}}}^{^{^{\! \! \! \! \! \! \! r_{ex}(\Theta)}}} 
 \sqrt{ \frac{2\mu}{\hbar^{2}} 
\ ( V_{TOT}(R,\Theta)-Q_{\alpha} ) } \ \  dr  \,\,,
\label{int}
\end{equation}

The reduced mass of $\alpha$ particle is denoted by $\mu$, $r_{en}(\Theta)$ and $r_{ex}(\Theta)$ 
are the classical
turning points determined for a given position of $\alpha$ particle by the relation 
$V_{TOT}(R,\Theta) = Q_{\alpha}$.
The energy released when the nucleus (Z,N) emits an $\alpha$ particle to the state $L+$
is obtained from the mass or total binding energies:
\begin{equation}
Q_{\alpha}(Z,N) = E_{B}(Z-2,N-2) - E_{B}(2,2) - E_{B}(Z,N) -E_{L+}(Z,N)\,\,,
\label{Qal}
\end{equation}
where $E_{L+}$ is the rotational energy of the $L+$ state of the  nucleus (Z,N).
The total potential ($V_{TOT}$) consists of Coulomb ($V_{C}$), nuclear ($V_{N}$) and centrifugal 
($V_{L}$) energies:
\begin{equation}
V_{TOT}(R,\Theta) = V_{C}(R,\Theta) + V_{N}(R,\Theta) + V_{L}(R,\Theta)\,\,,
\end{equation}
The centrifugal term of energy can be calculated:
\begin{equation}
V_{L}(R,\Theta) = \frac{\hbar^{2}  L(L+1)}{2\mu R(\Theta)^{2}}\,\,.
\end{equation}
The knocking frequency $\frac{\omega(\Theta)}{2\pi}$ is a slowly varying function of E.
\begin{equation}
\omega(\Theta) = 2 \pi \ \sqrt {\frac{E}{2 \ \mu \ r_{in}^{2}(\Theta)}}\,\,,
\end{equation}
In our investigations we accepted the  $\alpha$ particle kinetic energy $E = 112.1 \  MeV$ 
and $\alpha$ particle binding energies $E_{B}(2,2) = -28.30 \ MeV$
according to  \cite{PUDLINER,NOGGA}.
The obtained half-lives were averaged with the surface function in order to compare  them to the 
experimental data. 
This surface function in the axially symmetric geometry case is given by:
\begin{equation}
S = 2\pi \int_{0}^{\pi} dS(\Theta) = 2\pi \int_{0}^{\pi} \  d\Theta  \  R_{P}^{2}(\Theta) \ 
\sin(\Theta) \ \sqrt{1+\frac{1}{R_{P}^{2}(\Theta)}\left(\frac{dR_{P}}{d\Theta}\right)^{2}} \,\,,
\end{equation}
The mean half-life can be writen in the following way:  
\begin{equation}
\langle T_{1/2} \rangle = \frac{\int_{0}^{\pi} T_{1/2}(\Theta) \ 
 dS(\Theta)}{\int_{0}^{\pi} dS(\Theta)} \,\, .
\label{TSR}
\end{equation}

\section{Results}
As follows from the formulae presented in the previous chapter, the precise knowledge of the barriers
shapes depending on the angle $\Theta$  at which the particle is emitted from the nucleus is necessary
for the correct estimation of the tunnelling probability (formulae \ref{lmg},\ref{int}).
Figure 2 shown the obtained total barriers $V_{TOT}(R,\Theta)$ in the function of the distance between
the mass centres of $\alpha$ particle and the emitter for the deformed 
of the superheavy element $^{264}Hs_{156}$.
The equilibrium deformations \{$\beta_{2},\beta_{4},\beta_{6}$\} became taken from \cite{MOLLER95}.
The barriers for different orientations $\Theta$ of the
emitted $\alpha$ particle in the relation to the parent nucleus are shown. Note that position of the
barrier maximum and its value change with the angle of $\alpha$ cluster emission as shown in the upper
inside panel in figure 2. The lowest barrier height in the given example appears at the angle
$\Theta \approx 30^ { \circ }$ but the highest one is obtained for the equatorial position  
$\Theta \approx 90^ { \circ }$. The difference in the barrier maxima is  $\Delta V_{MAX} = 1.7 \  MeV$
for this nucleus. It has significant  influence on the half-lives as the difference in the barrier
height being only  $\Delta V_{MAX} = 0.5 \  MeV$ gives approximately the half-life differing
the order of magnitude.
  
Another parameter to which emission probability is very sensitive is the entrance energy of $\alpha$
particle in the barrier $V_{ENT}=V_{ST}+Q_{\alpha}$ where $Q_{\alpha}$ is the  
$\alpha$-decay energy calculated from the formula (\ref{Qal}) and is equal $Q_{\alpha} = 10.57 MeV$
for the studied isotope $^{264}Hs_{156}$. $V_{ST}$ is the energy of potential corresponding to the
configuration in the contact point between $\alpha$ particle and the parent nucleus ({\it sticking
configuration}). It is functional dependence on the angle $\Theta$ is shown in the bottom panel in
Figure 2. Taking into account the presented dependences leads to angular anisotropy and the largest
difference is  $\Delta V_{ENT} = 2.5  \  MeV$. As previously it can lead to essential differences in
the values of action integral, the obtained tunnelling probabilities and further in calculated
half-lives.

In order to show the effect of the deformation parameter $\beta_{6}$ on the obtained half-lives the
calculation for both cases $\beta_{6}>0$ and $\beta_{6}<0$ ($\vert\beta_{6}\vert=0.034$) was made 
leaving the deformation
parameters \{$\beta_{2},\beta_{4}$\} unchanged. Figure 3 show the course of variability of
tunneling probability function $I(\Theta)$  in both cases. Analyzing the diagrams it can be seen
that position of the curve maximum at $\beta_{6}<0$ denoted with the solid line 
correspond to the position of minimum for the curve at $\beta_{6}>0$  denoted with the broken line.
Position of the curve maximum for positive $\beta_{6}$ sign is shifted by about $15^{\circ}$
towards smaller angles compared to the curve minimum with negative sign of $\beta_{6}$.
As follows from the above analysis
half-lives are very sensitive to even smallest changes of deformation parameter $\beta_{6}$. 

On figure 4 calculated half-lives in the
function of angle $\Theta$ denoted by the thick solid line are shown.
The value $ \langle T_{1/2} \rangle$  obtained after averaging in
relation to the surface according to (\ref{TSR}) is also denoted (thin solid line). The arrow
indicates the experimental values \cite{ANT}. Half-lives differ from each other depending on the place of
$\alpha$ particle emission even by 5 orders of magnitude and the obtained mean value of half-life for
this element is close to the experimental value.

The dotted line correspond to the calculus without
taking centrifugal potential into consideration (emission from ground to the ground state is
considered). The calculated energy of the $2+$ state for this element is very small beeing   
$ E_{2+} = 46.6 \ keV $ see eg. \cite{MUNTIAN01}. One can also find that the contribution 
of the centrifugal barrier is small and it can be neglected. 

Systematic calculus of half-life for superheavy elements Z=108,110,112,114 is presented in figure 5
where decimal logarithms of the average half-life (dark filled squares) in the function of the mass
number A are shown. The obtained results are compared with literature calculations based on the
phenomenological formulae given by Viola and Seaborg (V-S) \cite{VIOLA,MOLLER97} 
denoted by crosses,
on the Generalized Liquid Drop Model (GLDM) \cite{TF} marked with blank squares and with available
experimental values (EXP) \cite{ANT} given by full triangles. The calculations were also made
not taking deformation into account denoted with open circles. As can been seen not taking
deformation into account leads to the lowest half-lives while including deformation degrees of
freedom increases significantly the values of half-life. The best agreement with the experimental
data is observed for the isotope with Z=110. For the element Z=108 the obtained results are the
closest to the predictions obtained using the Viola-Seaborg formulae \cite{VIOLA,MOLLER97} and for
the element Z=112 the largest agrement was obtained with calculations based on the GLDM model. 
The comparison of our results and those obtained from the Viola and Seaborg model shows that the
latter are systematicaly higher. 
The results for the isotope Z=114 obtained by us lie below the values obtained using
other models. Thus for the lately synthesized element $^{289}114$  the phenomenological
Viola-Seaborg formulae gives the value of half-life time logarithm  $log(T_{1/2})=4.86 \ s$ while
that calculated in this paper is $log(T_{1/2})=-1.30 \ s$ and the experimental data indicate 
$log(T_{1/2})=1.32 \ s$. For the neighbouring isotope $^{288}114$ the experimentally predicted
half-life is  $log(T_{1/2})= 0.26\ s$ see \cite{OGANESSIAN} ( and citations in it) but we obtain the
value almost three orders of magnitude smaller $log(T_{1/2})= -2.86 \ s$ and estimation based on
the Viola-Seaborg formulae gives the value $log(T_{1/2})= 2.80 \ s$ almost three times as large.
This can be explained by small equilibrium deformation of superheavy nuclei under discussion.
The results obtained for them are similar to those obtained for the spherical case 
and as mentioned earlier they are usually lower than the actual ones.
Despite some discrepancies the values of the half-time obtained by us not based on any 
phenomenological formula are similar to the experimental values.

\section{Conclusions}
The following conclusions can be drawn from our investigation devoted to $\alpha$-decay of
deformed superheavy elements:
\begin{itemize}
\item[1.] Taking deformation of nuclei into account affects significantly the potential
	  barrier shapes: position of maximum and entrance energies and in consequence, the
	  half-life time. 
\item[2.] Even small changes of deformation $\beta_{6}$ modify greatly the values of action
          integral. 
\item[3.] The half-life time for superheavy elements of  Z=$108 \div 114$ without 
          application of any phenomenological formula but taking into account 
	  deformation of nuclei in the ground state reproduce
          the experimental values well.
\item[4.] The centrifugal term affects the obtained results insignificantly.
                   
\end{itemize}

\newpage

\begin{center}
\begin{figure}

\psfig{figure=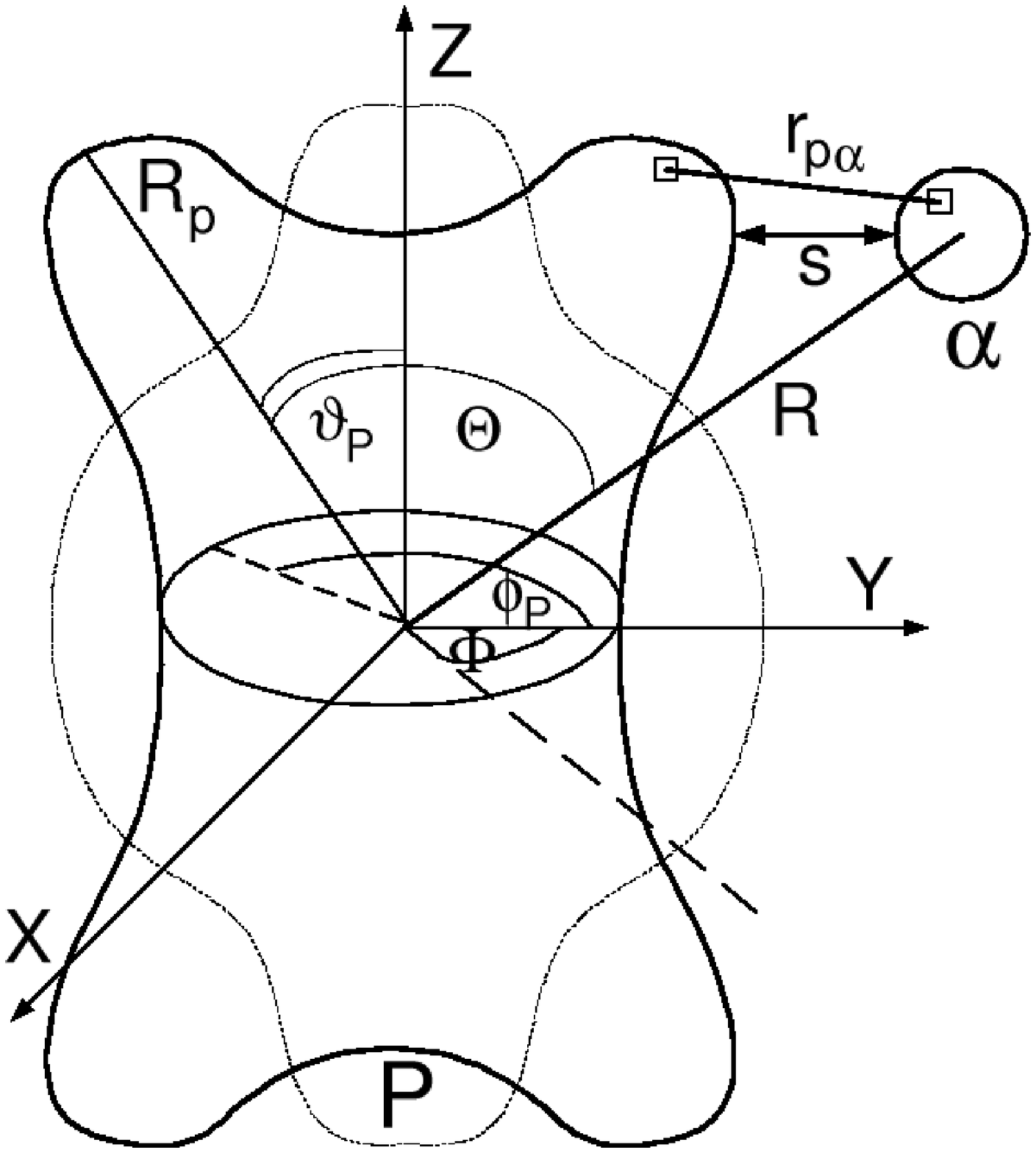,width=12cm,height=9cm}
\caption{An illustration of the geometry in the exit channel and 
the parameters emloyed for describing emission process.}

\end{figure}                                                                             
\end{center}

\begin{figure}
\psfig{figure=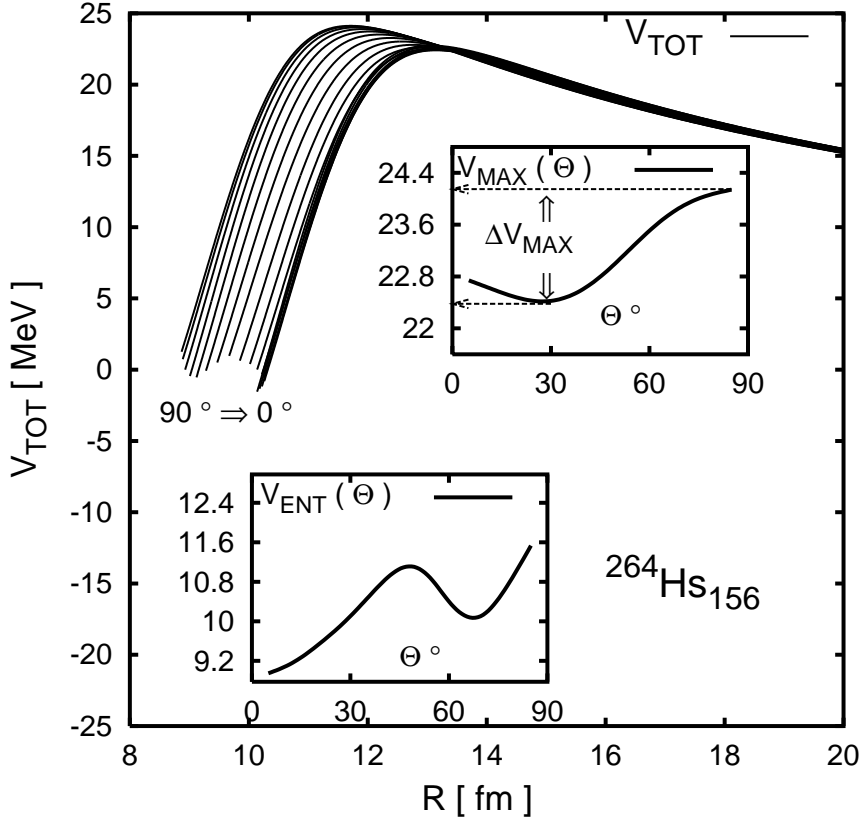}

\caption{The total potential ($V_{TOT}$) energy barriers obtained within the presented model 
(solid lines) dependent on the distance R  between the centers of the $\alpha$ particle
and the daughter nucleus for different orientations. The upper panel correspond to the  
maximum of the barrier ($V_{MAX}$) depending on $\Theta$, 
The bottom panel gives the entrance energy ($V_{ENT}$) during the emission process. 
All presented results are predicted for the exemplary deformed superheavy $ ^{264}Hs_{156}$ nucleus.}

\end{figure}                                                                    

\begin{center}
\begin{figure}
\psfig{figure=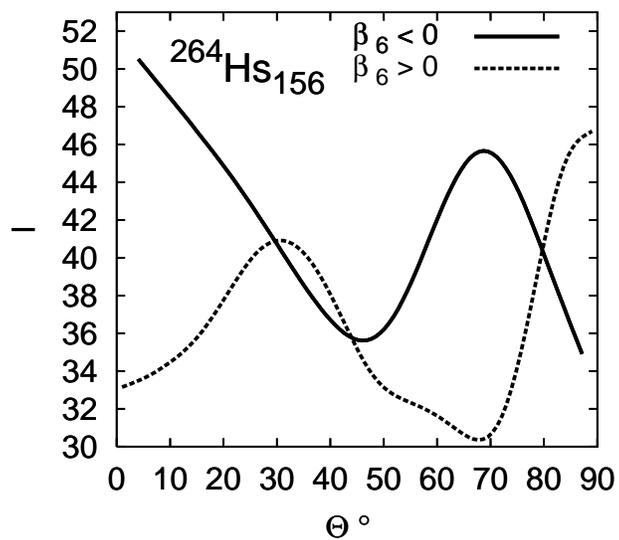}
\caption{The influence of emitted nucleus curvature on the obtained results is presented.
The penetration probability as a function of the position of $\alpha$ particle
for curvature with $\beta_{6} = -0.034$ (solid lines) and with 
$\beta_{6} = 0.034$ (dashed lines) for the deformed $ ^{264}Hs_{156}$ is shown. }
 
\end{figure}                                                                    
\end{center}

\begin{figure}
\psfig{figure=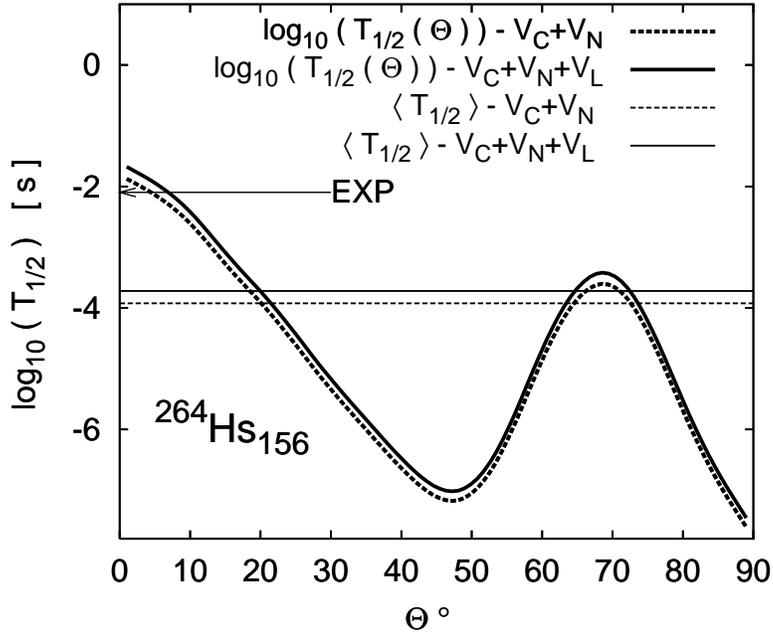}
\caption{The logarithm of the $\alpha$-decay half-life as an orientation function
(thick lines) is shown. Solid lines indicate that the centrifugal barier is included in calculation 
but the dotted ones
that it is not. The mean half-life $\langle T_{1/2} \rangle$ is marked by 
the thin lines. The solid line is related to the calculation with the centrifugal term and the 
 doted ones  to that without this term. Comparison with the experimental data (EXP) is  
indicated by the arrow \protect\cite{ANT}. }

\end{figure}

\begin{figure}
\psfig{figure=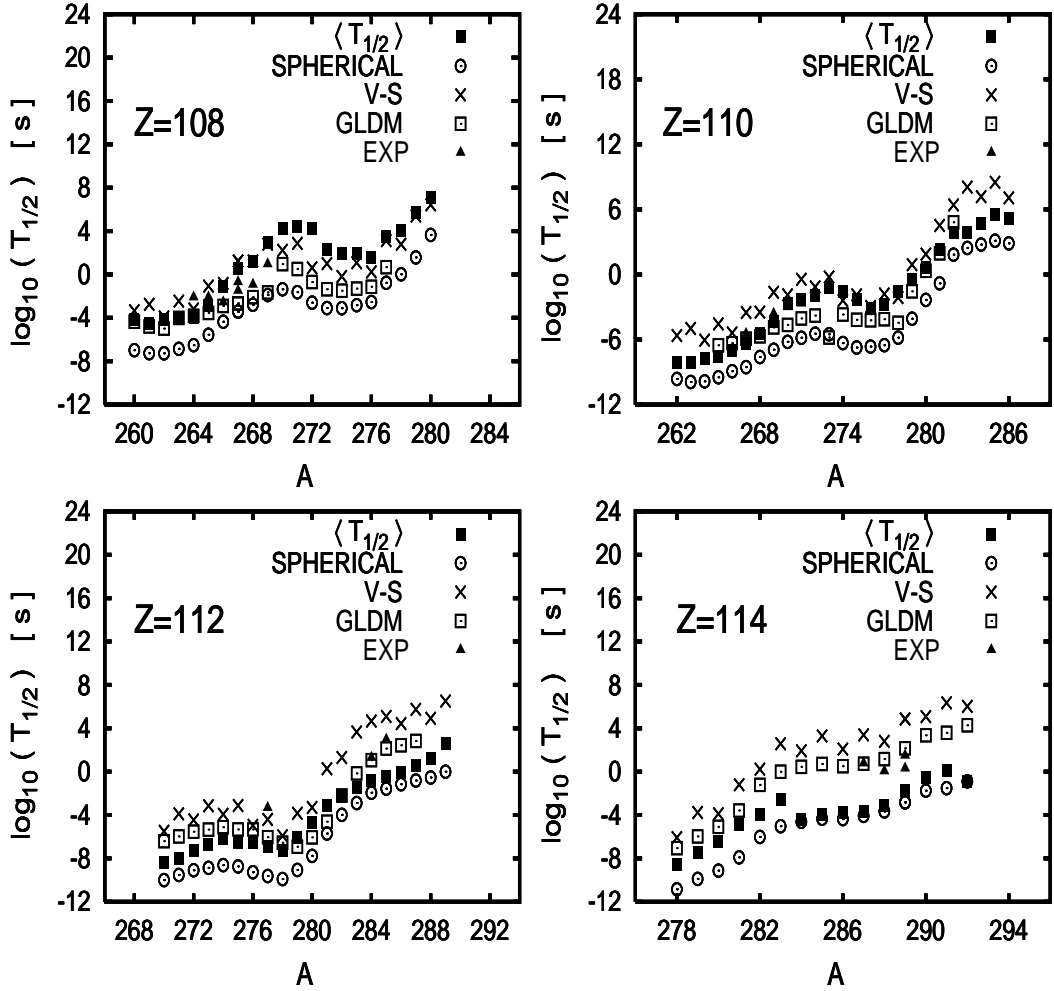,width=14cm,height=14cm}
\caption{ Logarithm of the $\alpha$-decay mean half-lives of the isotopes with atomic number 
Z = 108, 110, 112, 114 as a function of the mass number A. Full squares indicate the calculated
and averaged half-lives $\langle T_{1/2} \rangle$ . 
Open circles are related to our calculation without deformation (SPHERICAL), 
cross symbols denote the  calculation with the semi phenomenological Viola-Seaborg (V-S)
formula \protect\cite{VIOLA} with a new set of parameters given in \protect\cite{MOLLER97}.
The open squares were calculated using the GLDM model \protect\cite{TF}.
The triangles indicate the experimental data \protect\cite{ANT} (EXP).}

\end{figure}                                                                    

\end{document}